\documentclass [12pt]{article}
\usepackage{epsfig,amsmath,amssymb,graphicx}
\newcommand{\bs}{\begin{subequations}}
\newcommand{\es}{\end{subequations}}

\numberwithin{equation}{section}

\def \myfigures #1#2#3#4#5#6#7#8
{\begin{figure}[ht]
        \centering\includegraphics[width=#2 \textwidth]{#1.eps}
        \hfill
        \centering\includegraphics[width=#6 \textwidth]{#5.eps}
        \hfill
        \parbox[t]{#8\textwidth}{\caption {#7}\label{#5}}
\end{figure} }

\title{Vortex structure in exponentially shaped Josephson junctions}

\bigskip

\author{Yu.M. Shukrinov$^{1,2}$, E.G. Semerdjieva$^{1,3}$ and T.L. Boyadjiev$^1$
\\
{\it \small
$^1$Joint Institute for Nuclear Research, Dubna, Russia,141980
}\\
{\it \small $^2$Physical Technical Institute, 299/1 Aini Str., Dushanbe,
Tajikistan,734063
}\\
{\it \small $^3$Plovdiv University, 24 Tzar Assen Str., Plovdiv, Bulgaria, 4000
}\\
{\small e-mail: {\it $^2$shukrinv@thsun1.jinr.ru }}}

\date{}

\begin{document}

\maketitle

\begin{abstract}

We report the numerical calculations of the static vortex structure and critical curves
in exponentially shaped long Josephson junctions for in-line and overlap geometries. Each
solution of the corresponding boundary value problem  is associated with the
Sturm-Liouville problem whose minimal eigenvalue allows to make a conclusion about the
stability of the vortex. The change in width of the junction leads to the renormalization
of the magnetic flux in comparison to the case of a linear one-dimensional model. We
study the influence of the model's parameters and, particularly, the shape parameter on
the stability of the states of the magnetic flux. We compare the vortex structure and
critical curves for the in-line and overlap geometries. Our numerically constructed
critical curve of the Josephson junction matches well with the experimental one.

\end{abstract}

\section*{Introduction}

The Physical properties of the magnetic flux in Josephson junctions are the base of modern superconducting electronics. The shape of the junction has an essential influence on its properties \cite{barone}. One of the very interesting objects is the exponentially shaped Josephson junction (ESJJ). The unidirectional flow of the magnetic flux quanta in the long Josephson junction provides a means for tunable oscillators at the frequencies above 100 GHz with power output of a few microwatts. The ESJJ decrease the spectral linewidths of the oscillator, which is large in junctions with rectangular shape. The exponential shape provides better impedance matching to an output load and allows to avoid a chaotic regime that exists for rectangular junctions \cite{bcs_96}.

Josephson junctions with an exponentially varying width in the $xy$ plane have been
studied theoretically and experimentally in \cite{bcs_96,cmc_02}, where the influence of
the junction's shape on the current--voltage characteristics of the junctions was
studied in detail. However, the problems involving the determination of the stability
regions of the static distributions and the structure of the critical curves have not
been adequately studied. The present paper is devoted to a study of those questions,
which are important from the applied and theoretical points of view.

\section{Method}

The basic equation for the phase $\varphi(t, x)$ in the ESJJ  can be written in the form
\cite{bcs_96} -- \cite{gsk_00}
\begin{equation} \label{time}
    \ddot \varphi + \alpha \dot\varphi - \varphi\,'' + \sin \varphi + g(t,x) + \gamma = 0\,.
\end{equation}
Here $x \in (0,l)$ and $t>0$. The spatial coordinate $x$ is normalized to the Josephson penetration depth, $l$ is the dimensionless junction's length, $\gamma$ is external current and time $t$ is normalized to the plasma frequency \cite{barone}. An overdot denotes differentiation with respect to time $t$, and a prime denotes differentiation with respect to the spatial coordinate $x$. The parameter $\alpha$ describes the dissipative effects. The term
\begin{equation}\label{geom}
    g(t,x) \equiv \sigma \left[ \varphi\,'(t,x) - h_B \right]
\end{equation}
in the equation \eqref{time} describes the additional current in the exponentially shaped
Josephson junctions with the shape parameter $\sigma > 0$. As it is known
\cite{barone}, by appropriate normalization we can interpret $\varphi(t,x)$ as
dimensionless magnetic flux along the junction. Then $h_B$ represents the dimension\-less intensity of the external (boundary) magnetic field.

In this paper we consider the static distributions $\varphi(x)$ of the magnetic flux
along the junction. So, the corresponding boundary value problems (BVP's) are posed as
follows
\begin{itemize}
    \item[$\surd$] for the in-line geometry
        \bs \label{stati}
        \begin{gather}
            -\varphi\,'' + \sin \varphi  + g(x) = 0\,, \label{steqi}\\
            \varphi\,'\left( 0 \right) - h_B  + l\gamma  = 0, \label{bcai} \\
            \varphi\,'\left( l \right) - h_B  = 0\,, \label{bcbi}
        \end{gather}
        \es
  \item[$\surd$] for the overlap geometry
        \bs\label{stato}
        \begin{gather}
            -\varphi\,'' + \sin \varphi  + g(x) + \gamma = 0\,, \label{steqo}\\
            \varphi\,'(0) - h_B = 0, \label{bcao} \\
            \varphi\,'(l) - h_B = 0\,, \label{bcbo}
        \end{gather}
        \es
\end{itemize}

The solutions of the BVP's  \eqref{stati} and \eqref{stato} depend on the set of four
parameters $p\equiv\{l,\sigma, h_B, \gamma\}$  of the model, i.e. $\varphi = \varphi(x,p)$.

In order to study the stability of the solutions $\varphi(x)$, following \cite{galfil_84} we pose \cite{sbs_04} the Sturm-Liouville problem (SLP)
\bs\label{c3}
    \begin{gather}
         -\psi\,'' + \sigma \psi\,' + q\,(x)\, \psi  = \lambda \,\psi \,,\quad x\in(0,l)\,, \label{slp} \\
        \psi\,'(0) = 0, \quad \psi\,'(l) = 0\,,
    \end{gather}
\es Here $\lambda$ is the spectral parameter. The potential $q(x) \overset{def}{=} \cos \varphi(x)$ corresponds to the concrete static solution $\varphi(x)$.

We note that the function $q(x,p)$ is limited $|q(x,p)|\leq 1$ on the finite interval $[0,l]$ for every $p \in \mathcal{P} \subset \mathbb{R}^4$. Then the SLP \eqref{c3} have \cite{ls_88} bounded below counted sequence of real eigenvalues (EV) $\{\lambda_n\}$, $n=0,1,\ldots$ Every EV $\lambda_n$ corresponds to an unique real eigenfunction (EF) $\psi_n(x)$, which satisfies the norm condition
\begin{equation} \label{norm}
    \int\limits_0^l\psi^2_n(x)\,dx = 1\,.
\end{equation}
The number of zeroes of EF  $\psi_n(x)$ in the interval  $(0,l)$ is equal to $n$. Particularly, the EF $\psi_0(x)$, which corresponds to the minimal EV $\lambda_{min}\equiv \lambda_0$ does not have zeroes for $x \in (0,l)$.

Because the solution  $\varphi(x,p)$ depends on the parameters $p$, the potential of SLP induced by this solution and the corresponding EV and EF depend on the parameters as well, i.e., $\lambda_n=\lambda_n(p)$  and $\psi_n=\psi_n(x,p)$.

The static solution is exponentially stable \cite{galfil_84, sbs_04} in some subset of the parameter region $\mathcal{P}$ if the minimal EV satisfies $\lambda_{min}(p) > 0$. When $\lambda_{min}(p) < 0$ the solution $\varphi(x)$ is unstable. The points $p\in\mathcal{P}$  lying in the hypersurface
\begin{equation}\label{biff}
    \lambda_{min}(p) = 0
\end{equation}
are the bifurcation points for the solution $\varphi(x,p)$. The values of the parameters that satisfy \eqref{biff} are called critical (bifurcation) values for the solution  $\varphi(x)$. The cross section of the surface \eqref{biff} by hyperplane corresponding to fixed values of two parameters determines the critical curve for another two parameters. From an experimental point of view, the most important are the critical curves ``current-magnetic field''
\begin{equation}\label{gamhb}
    \lambda_{min}(\gamma, h_B) = 0\,,
\end{equation}
for fixed length $l$ and shape parameter $\sigma$.

We have to note that mathematically it is possible to pose two different problems for the equations \eqref{stati} -- \eqref{norm}.

Let $\varphi(x)$ be a solution of BVP \eqref{stati} or \eqref{stato} corresponding to a fixed set of parameters $p$. We can  check its stability by means of SLP \eqref{c3}, \eqref{norm}. Let us suppose that $\lambda_{min}(\gamma, h_B) > 0$. In order to obtain a point from the critical curve \eqref{gamhb}, we have to change $h_B$ for given $\gamma$
(or change $\gamma$ for given $h_B$) until the condition \eqref{gamhb} is fulfilled.

In order to calculate bifurcation points directly, we have to solve  the system \eqref{stati} or \eqref{stato}, \eqref{c3} and \eqref{norm} for given $\lambda_{min}$ as a non-linear eigenvalue problem with respect to the couple $\{\varphi(x), \psi(x)\}$ and one of the parameters $p$\ \cite{fgbp_pla87} -- \cite{bpp_nma88}. Such an approach has been successfully applied to different nonlinear physical problems \cite{tlb_02}.

In this paper, we find the regions of stability for distributions of magnetic flux in ESJJ
by numerical calculations. Let us note that the replacement of the problem for stability
of the solutions of nonlinear operator equations by the eigenvalue problem for linear
operator has the rigorous mathematical explanation \cite{kra_54}, \cite{ka_74}.

\section{Results and discussion}

The curves of $\lambda _{\rm min}(h_{B})$ for the Meisner distribution and the first few
stable vortices in a Josephson junction in the in-line geometry for current $\gamma
 = 0$ and two values of shape parameter $\sigma  = 0$ and $\sigma = 0.07$ are demonstrated in Fig.\ \ref{fig2}.
\myfigures{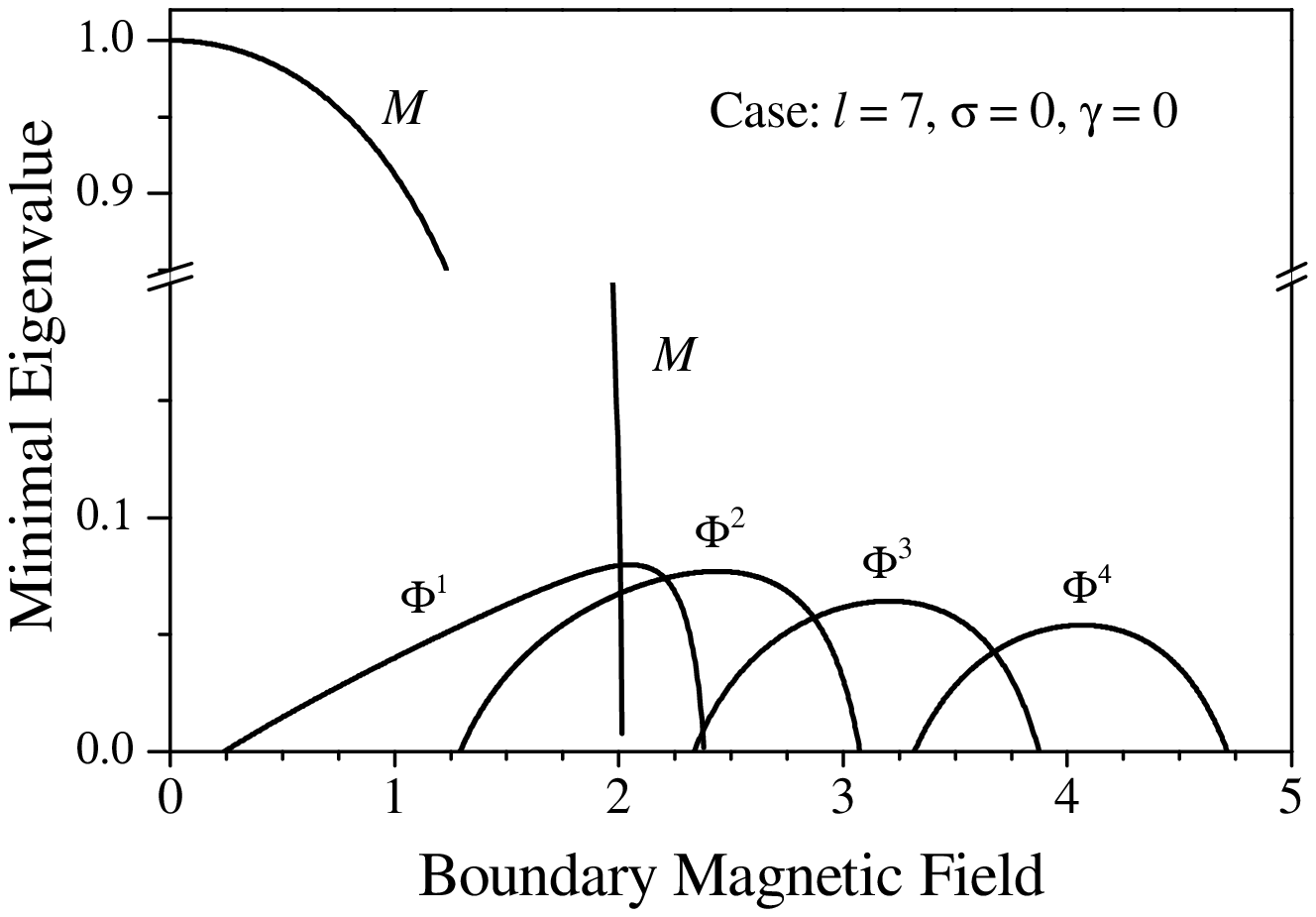}{0.43}{}{0.47} {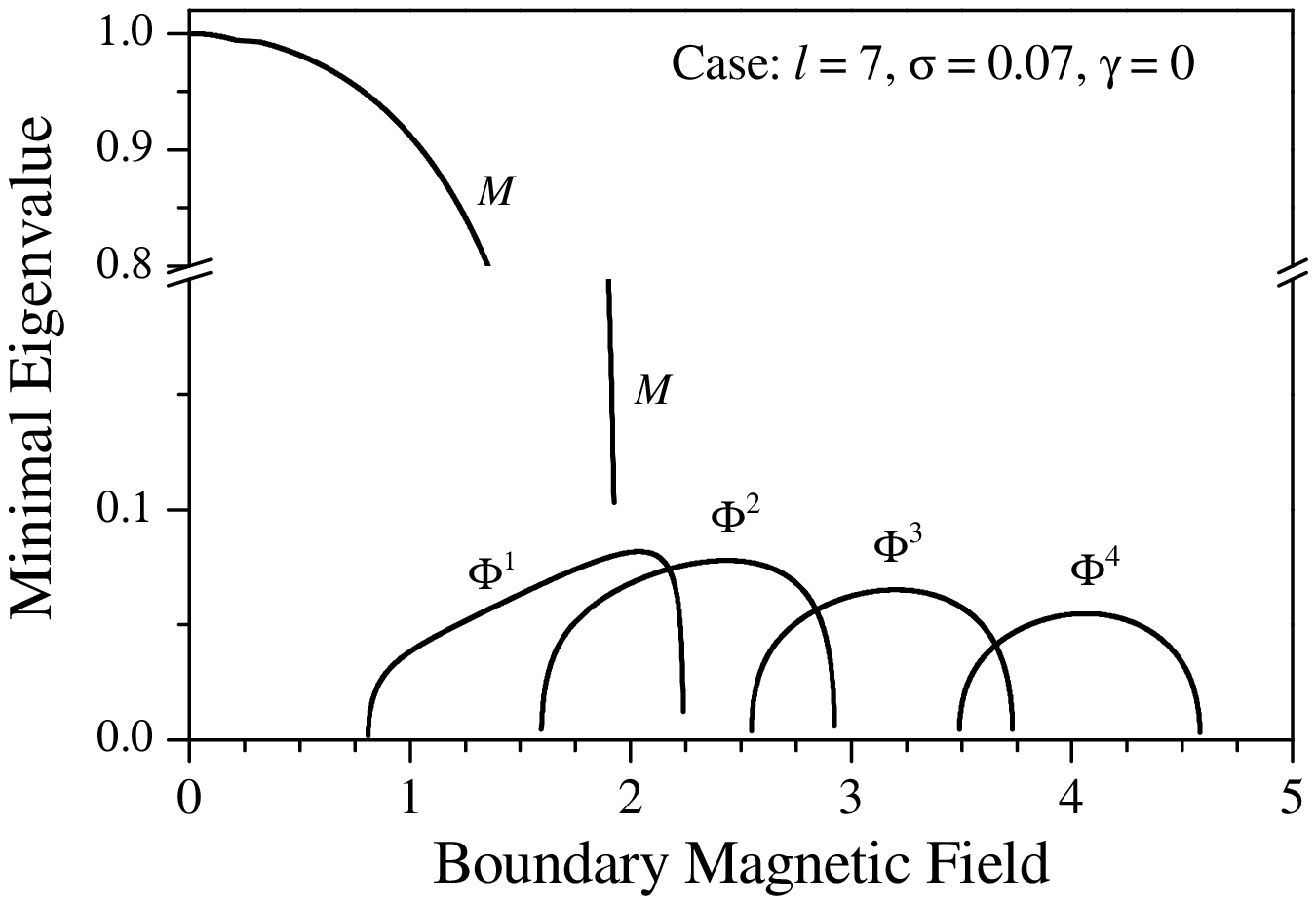}{0.43}{The $\lambda_{\rm min}(h_{B})$ curves for different $\sigma $}{0.51}

Each curve has two zeroes, the distance between which determines the stability region of
the vortex upon variation of the magnetic field $h_{B}$. The zeroes themselves are
critical values of the field $h_{B}$ for zero current $\gamma $. It is well seen that the
role of the shape parameter $\sigma $ is most important for small values of $h_{B}$. In
particular, when $\sigma  = 0$ the $\Phi^{1}$ vortex exists already for $h_{B}\approx
0.054$. For $\sigma  = 0.07$, however, the existence region in respect to field $h_{B}$
is considerably compressed, and the vortex exists starting for $h_{B}>0.75$. The compression of the $\lambda _{\rm min}(h_{B})$ curves for different vortices decreases rapidly with increasing $h_{B}$.

The dependencies of the minimal eigenvalue of the SLP for the fluxon $\Phi ^{1}$
on the shape parameter $\sigma $ in case of in-line geometry for $h_B=1$ and $h_B=2$ are shown
in Fig.\ \ref{fig4}. It is seen that for fixed $h_{B}$ and $\gamma $ there exists a
certain maximum value of $\sigma $ above which the distribution of $\Phi ^{1}$ loses
its stability, i.e., a bifurcation of the vortex occurs upon a change in $\sigma $. Large
values of the magnetic field $h_{B}$ correspond to large critical values of $\sigma $.

\myfigures{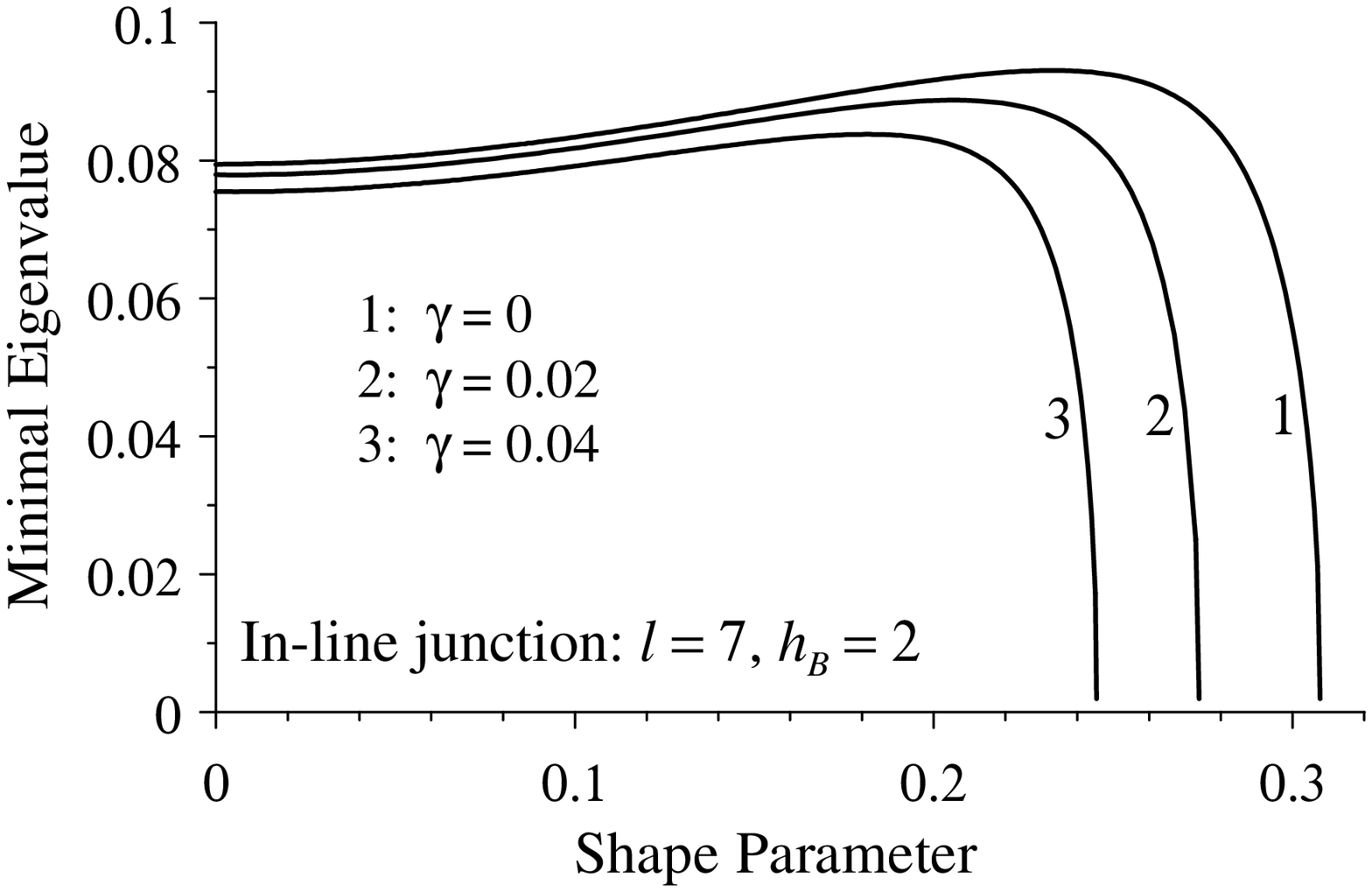}{0.44}{}{0.5}{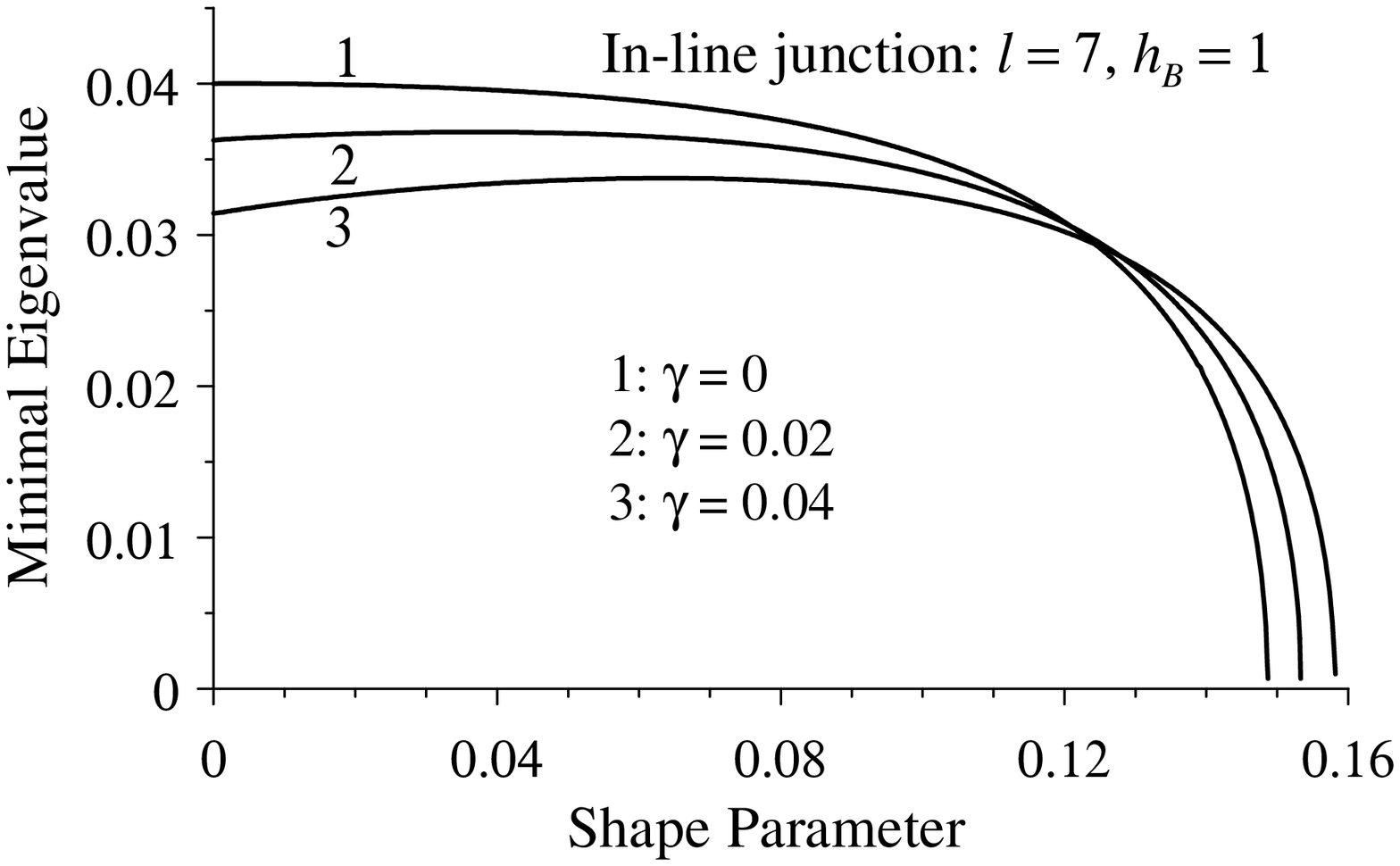}{0.45} {Bifurcation upon a change in
$\sigma $ for $h_B=1$ (left) and $h_B=2$ (right) in-line geometry}{1}

The influence of the external magnetic field $h_{B}$ on the magnetic flux distribution
$\varphi\,'(x)$ in the in-line geometry for the fluxon $\Phi ^{1}$ for $\sigma  = 0.07$ was
studied in \cite{sbs_04}. For a certain value $h_B = h_m \approx 1.515$, the maximum of
the derivative $\varphi\,'(x)$ is localized in the middle of the junction. For
$h_{B} < h_{m}$ the fluxon is ``expelled'' to the end $x = l$ by the ``geometric'' current
$g(x)$. If the length of the Josephson junction is sufficiently large, then a change of
the current $\gamma $, equivalent to a change in magnetic field for the left end $x = 0$,
turns out to have only a slight influence on the local maximum magnetic field in the
junction. For $\sigma  = 0$ the variation of the current causes the maximum of the magnetic
field to shift to the right or left of the center of the Josephson junction, depending
on the direction of the current. For values $h_{B}>h_{m}$ the scheme $x_{m}$ of the maximum of the field $\varphi '(x)$ of the fluxon $\Phi ^{1}$ is shifted to the left from the center $x = l/2$ toward the end $x = 0$.

As we can see in Fig.\ \ref{fig4}, the stability region in case of in-line geometry is
increased for $h_B=1$ ($h_B < h_m$) and decreased for $h_B=2$ ($h_B > h_m$) with the
increase of the external current.

For the overlap geometry (see Fig.\ \ref{fig6}.) we have the opposite situation. The stability region is decreased for $h_B=1$ ($h_B < h_m$) and increased for $h_B=2$ ($h_B > h_m$) with the increase of the external current.
\myfigures{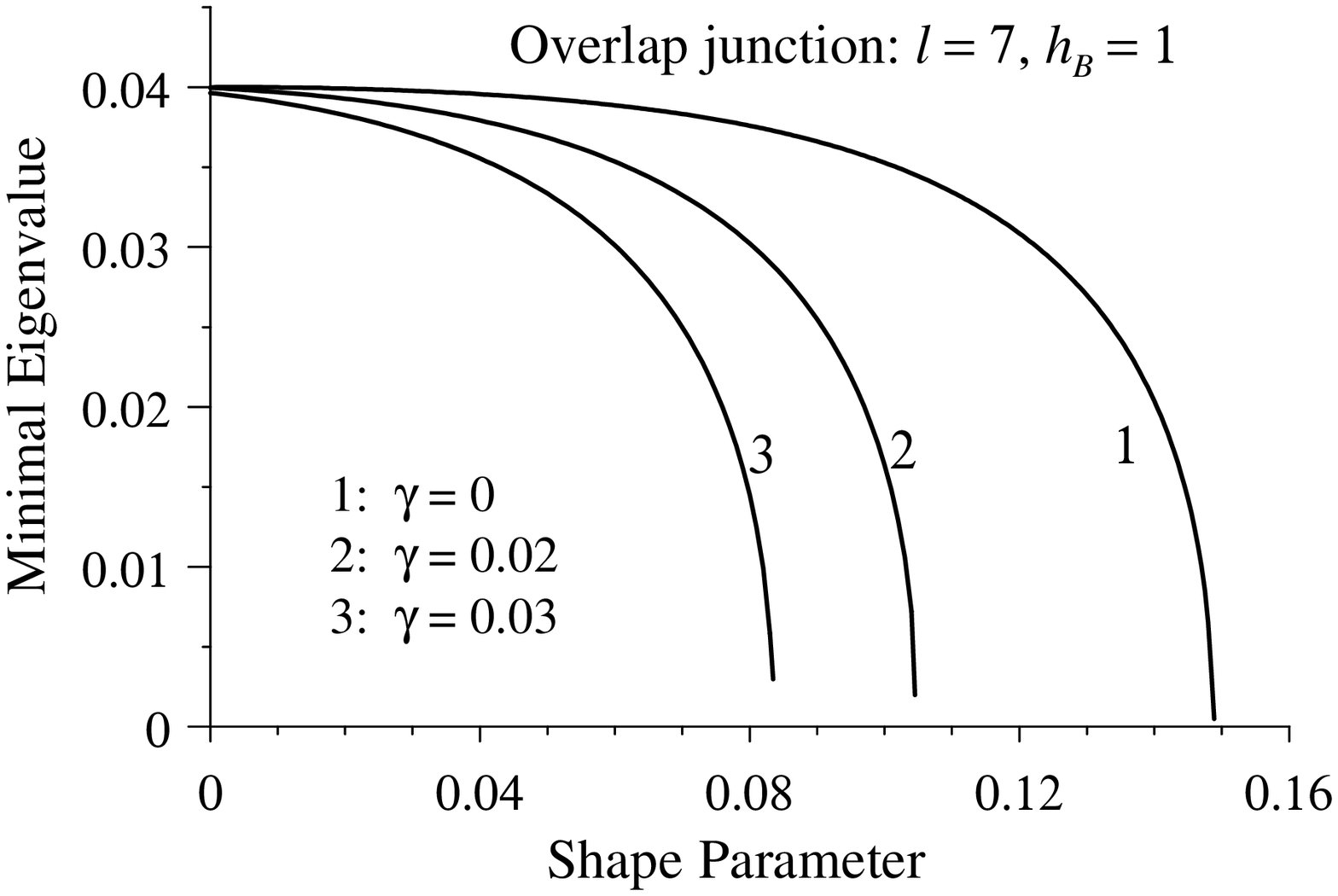}{0.44}{}{0.5} {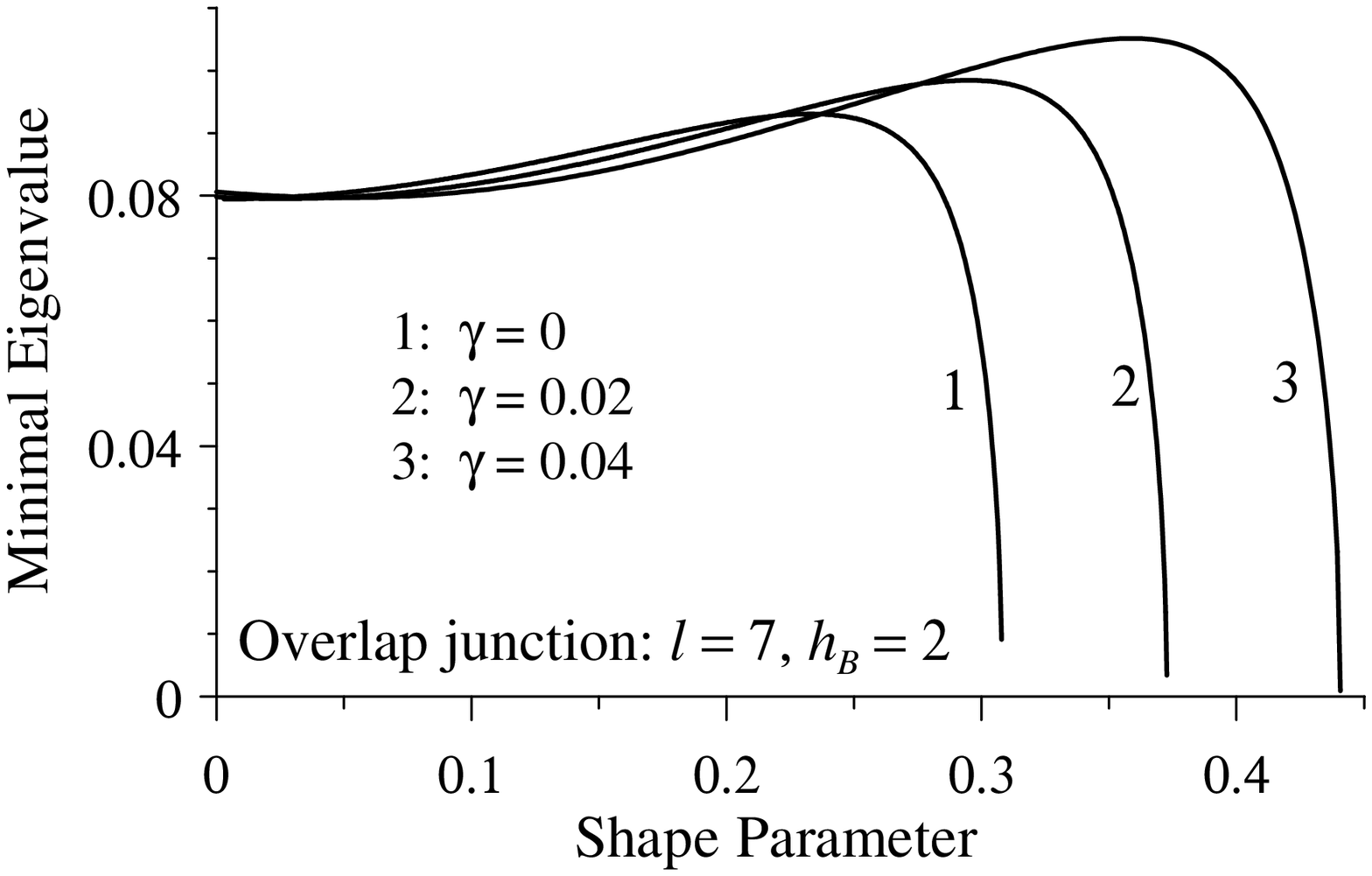}{0.45} {Bifurcation upon a change in $\sigma $ for $h_B=2$ (left) and $h_B=2$ (right) in overlap geometry}{1}

When $\gamma = 0$ the BVPs for the the in-line and overlap geometries coincide. In case of in-line geometry the increase of $\gamma$ leads to the decrease of $h_B$ at the end $x=0$. The sign of $g(x)$ depends on the value of $h_B$ and for $h_B>h_m$ its sign is negative. When $h_B<h_m$, there is a region on $x$ where $g(x)>0$. In the first case, the BVP \eqref{stati} leads to the increase of the stability region with the increase of $\gamma$, in the second one -- to its decrease. In the overlap geometry the boundary conditions do not change, but the additional term $\gamma$ in the BVP \eqref{stato} effectively changes the absolute value of $g(x)$. For $g(x)<0$ ($h_B>h_m$) the stability region is increased with $\gamma$, because $\gamma$ decreases the absolute value of $g(x)$. When $h_B<h_m$ external current increases the interval in which $g(x)>0$ and this leads to the decrease of the stability regions as shown in Fig.\ \ref{fig6} (left).
\myfigures{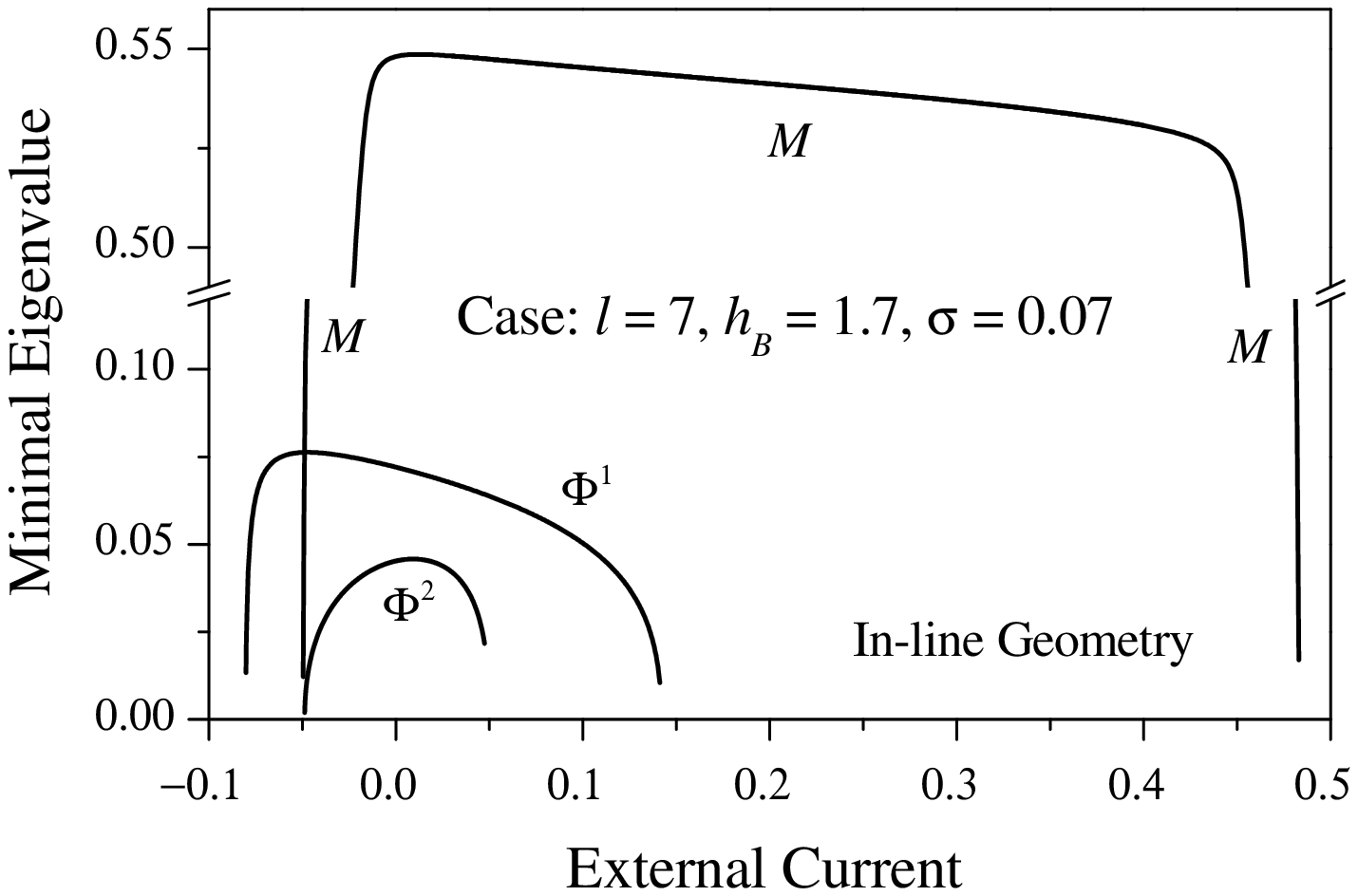}{0.45}{}{0.5} {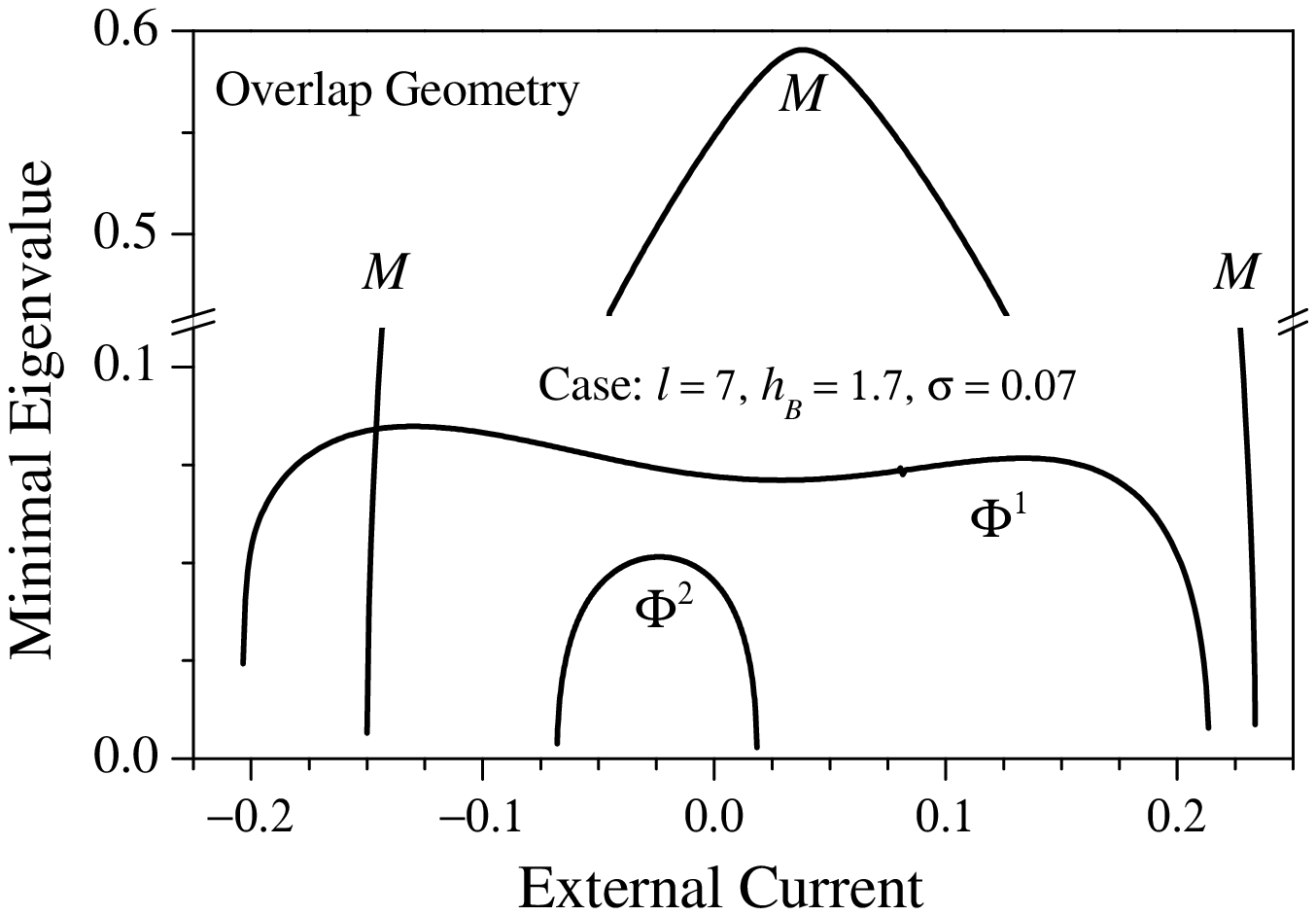}{0.45} {The $\lambda _{\rm min}(\gamma)$ curves
for the in-line (left) and overlap (right) geometries}{1}

Let us now consider the question of constructing by numerical means the critical current
versus magnetic field relation, which is determined implicitly by Eq.\ \eqref{gamhb} for
each magnetic flux distribution in the Josephson junction. The importance of this problem
stems from the possibility of measuring this relation
experimentally \cite{cmc_02, galfil_84, changho_84, vdks_88, lmu_94, lmu_94a}. We note that for different configurations of the magnetic flux in the Josephson junction the values of the critical parameters (in particular, the critical current and magnetic field) can be substantially different. Therefore we have to distinguish the critical parameters for the specific distributions and for the Josephson junction as a whole.

Figure \ref{fig8} shows the $\lambda_{\rm min}(\gamma )$ curves for stable solutions of
the BVP \eqref{stati} in a field $h_{B} = 1.7$ for both in-line and overlap geometries. The
distances between zeroes of the functions are the stability intervals of the
corresponding distributions with respect to the current $\gamma $. The right and left
zeroes of the function $\lambda _{\rm min}(\gamma )$ are the positive and negative
critical currents, respectively, of the distribution in the given field $h_{B}$. Because
of the asymmetry of the boundary conditions in in-line geometry for $\gamma >0$ the
critical current of the Meissner distribution [which we denote by $\gamma _{c}(M)$] is
the highest, but for $\gamma <0$ the largest in module is the critical current $\gamma
_{c}(\Phi ^{1})$ of the vortex $\Phi ^{1}$. Consequently, in a field $h_{B} = 1.7$ the
positive critical current of the junction is $\gamma _{c} = \gamma _{c}(M)$, while the
negative critical current is $\gamma _{c} = \gamma _{c}(\Phi ^{1})$.
\myfigures{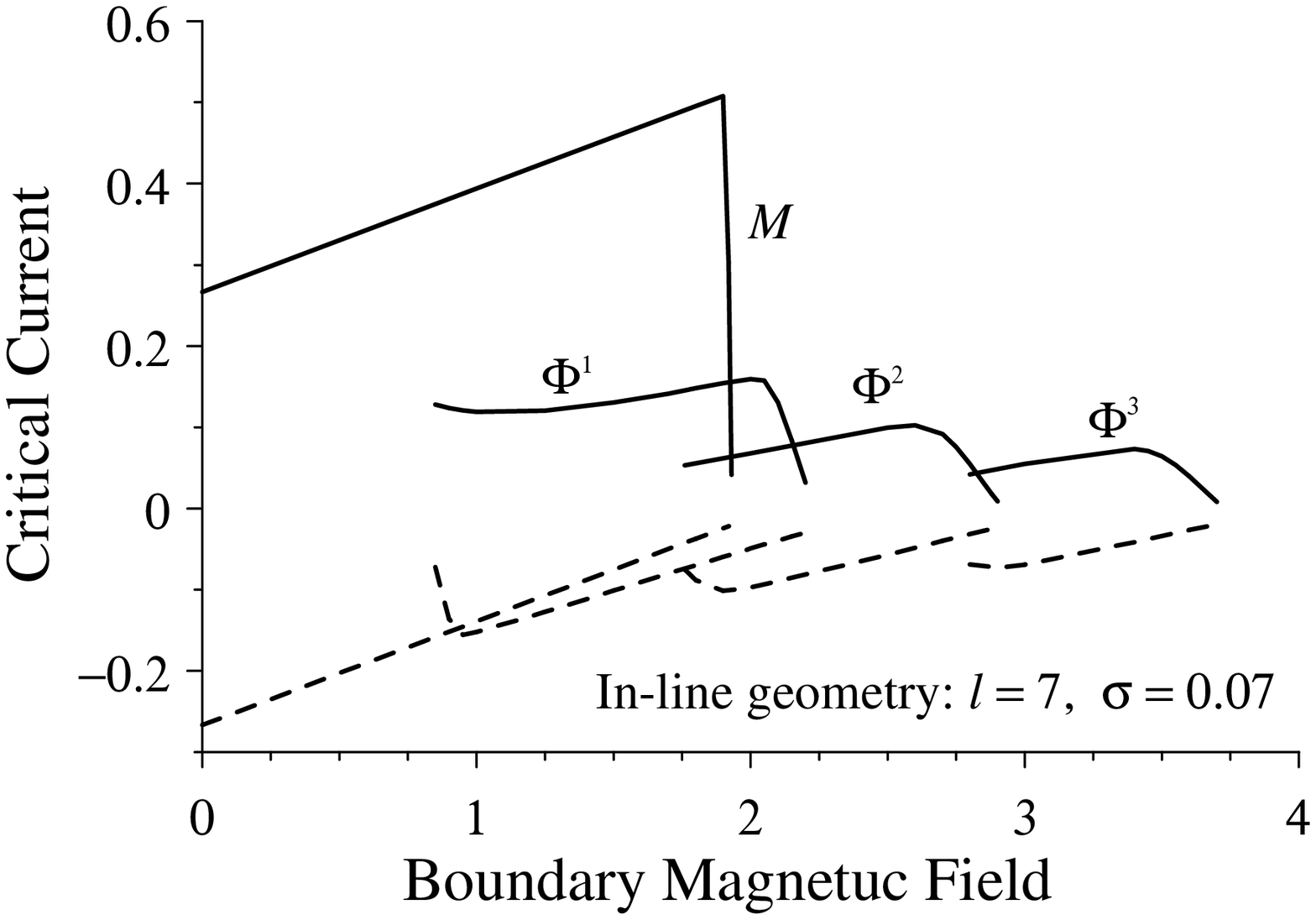}{0.45}{}{0.5} {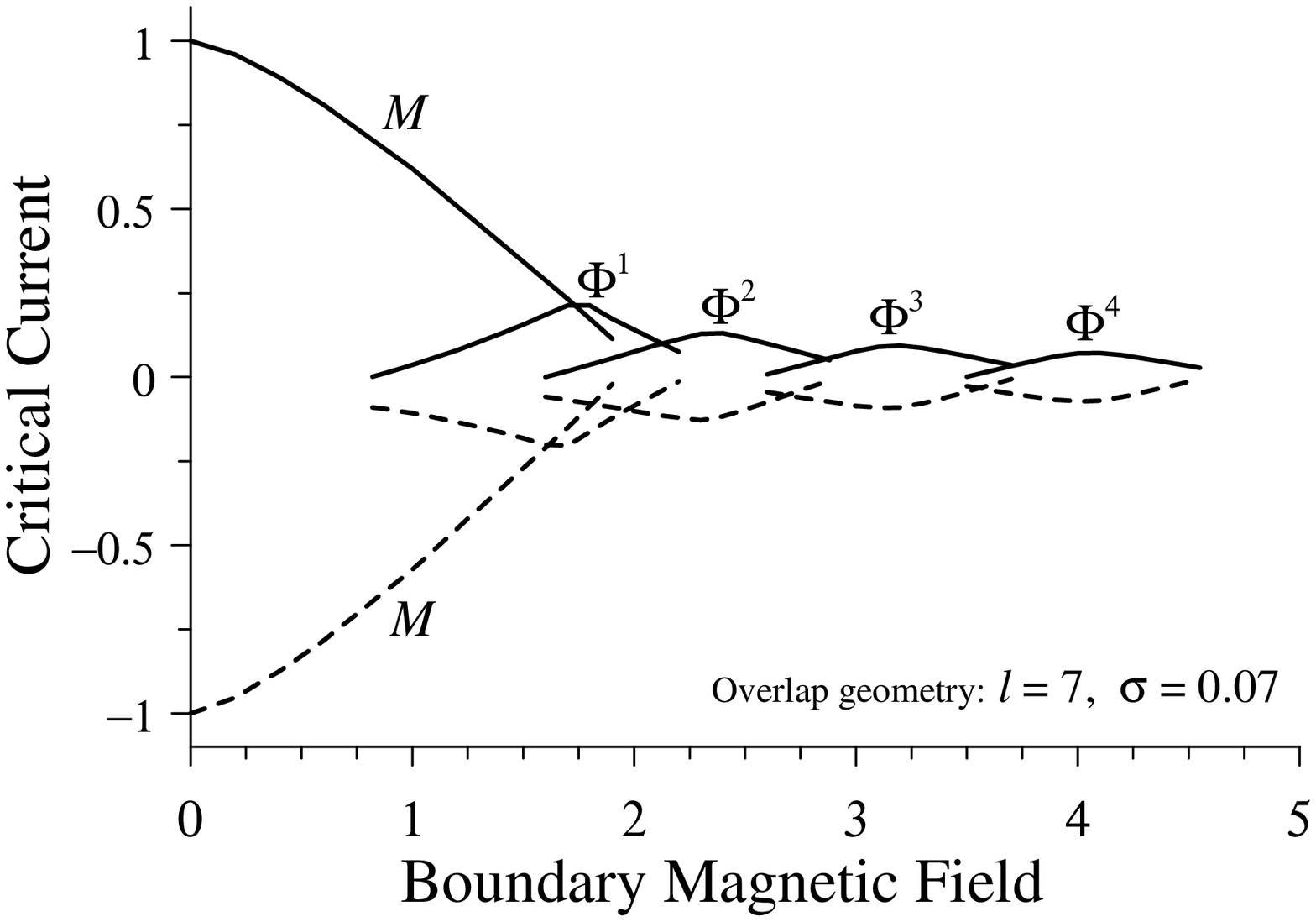}{0.45} {The critical curves of the junction for
the in-line (left) and overlap (right) geometries}{1}

We note that for the in-line geometry the curve for the $M$ distribution has a
characteristic plateau --- the ``breakoff'' of the Meissner solution sets in for a rather large module of the external current.

The critical curve $\gamma _{c}(h_{B})$ for a junction is constructed as the envelope of the critical curves corresponding to different magnetic flux distributions in the junction. In other words, the critical curve consists of pieces of the critical curves for individual states with the largest module of the critical current for a given $h_{B}$. The parts of the critical curves corresponding to the intervals $h_{B}\in[0,4)$ for in-line and $h_{B}\in[0,5)$ for overlap geometries are illustrated in Fig.\ \ref{fig10}. We note that our numerically constructed critical curve of the Josephson junction for the in-line geometry matches well with the theoretical and experimental results presented in Figs.\ 6a and 7a of Ref. \cite{cmc_02}. Critical curves in case of overlap ESJJ's have not been obtained experimentally yet.

\section{Conclusions}

Numerical calculations of the magnetic flux distributions and their bifurcation with parameters of the model in the long ESJJ have been done. To study the stability of the solutions, we associate with each distribution the SLP whose potential is defined by this distribution. We showed that the magnetic flux can change its stability for some critical values of the parameters. The different behaviour of the bifurcation upon a change in the shape parameter for different values of the boundary magnetic field and the external current was investigated. The envelope curve of the critical curves for several fluxon states was obtained as the critical curve for the junction in cases of in-line and overlap geometries. Our numerically constructed critical curve of the exponentially shaped Josephson junction matches well with the experimental data.

\subsection*{Acknowledgements}

Yu.M.S. gratefully acknowledges the financial support of INTAS grant No. 01-0617.


\end{document}